# AN ANALYSIS OF THE EXPECTED DEGRADATION OF SILICON DETECTORS IN THE FUTURE ULTRA HIGH ENERGY FACILITIES[&,*]

IONEL LAZANU
*University of Bucharest, Faculty of Physics, POBox MG-11, Bucharest-Magurele, Romania*
*e-mail: ionel.lazanu@gmail.com*

SORINA LAZANU
*National Institute for Materials Physics, POBox MG-7, Bucharest-Magurele, Romania*

**Abstract**

In this contribution we discuss how to prepare some possible detectors – only silicon option being considered, for the new era of HEP challenges because the bulk displacement damage in the detector, consequence of irradiation, produces effects at the device level that limit their long time utilisation, increasing the leakage current and the depletion potential, eventually up to breakdown, and thus affecting the lifetime of detector systems. Physical phenomena that conduce to the degradation of the detector are analysed both at the material and device levels, and some predictions of the time degradation of silicon detectors in the radiation environments expected in the LHC machine upgrade in luminosity and energy as SLHC or VLHC, or at ULHC are given. Possible effects at the detector level after high energy cosmic proton bombardment are investigated as well. Time dependences of these device parameters are studied in conditions of continuous irradiation and the technological options for detector materials are discussed, to obtain devices harder to radiation.

## 1. Introduction: the Future Ultra High Proton Facilities

Particle physics has made striking advances in the last fifty years in describing the intimate structure of matter and the forces that determine the architecture of the universe. Major steps towards answering the open questions would come from developing global theories but invoke

---

[&] Work in the frame of the CERN RD-50 Collaboration

[*] This work has partially been supported by the Romanian Scientific National Programmes CERES and MATNANTECH, under contracts C4-69/2004 and 219 (404) /2004 respectively.

energy scales that are far beyond the reach of conceivable accelerators and from obtaining new experimental results at higher energies, or at least observable traces of these theories at accessible energies. Experiments relevant to this quest are not restricted to the highest energy frontier; they are also possible at low energy scales. A discussion of the search for new physics at the future collider facilities is given in Ref. [1].

The LHC at CERN will be operational from 2007. But the physics results expected from this facility will not be able to clarify all the unknowns of the Standard Model. So, despite the technological difficulty, significant upgrades of the accelerator in energy and luminosity are considered as SLHC and VLHC respectively. The upgrade path will be defined by the results from the initial years of LHC operation, and now only possible scenarios can be done up to 240 TeV, the final energy of the project. Ultimate Large Hadron Collider (ULHC) is imagined as accelerating protons to energies above $10^7$ TeV in a ring with a circumference L≈4×$10^4$ km around Earth and such a facility will permit to ex-plore the "intermediate" energy scale, about $10^7$ TeV, between the electroweek scale (1 TeV) and the scale of grand unification (around $10^{14}$ TeV).

Table 1 Main parameters of future colliders

| Parameters | LHC | SLH (stage 1) | SLHC (stage 2) | VLHC (stage 1) | VLHC (stage 2) | ULHC |
|---|---|---|---|---|---|---|
| √s [TeV] | 14 | 14 | 28 | 40 | 175 (up to 240) | 2×$10^7$ |
| Circumference [km] | 26.7 | 26.7 | 26.7 | 233 | 233 | 40000 |
| Luminosity ×$10^{34}$ [$cm^{-2}s^{-1}$] | 1 | 10 | 1 (10) | 1 | 1÷10 | ? 1 (10) |
| Bunch spacing [ns] | 25 | 12.5 | | 18.8 | 18.8 | |
| $\sigma_{pp}$-total [mb] | ≈ 117 | | ≈ 136 | ≈ 143 | ≈ 166 | 770 |
| $\sigma_{pp}$-inel. [mb] | ≈ 80 | | ≈ 90 | ≈ 100 | ≈ 140 | ≈ 270 |
| Interactions | 20 | 100 | | | up to 250 | |
| <$E_T$> charged part.[MeV] | 450 | | 500 | | 600 | 1280 |

Other possibility to have access to ultra high energy particles, complementary to colliders is the exploitation of the astroparticles or cosmic rays (CR). Protons are the most abundant charged particles in space and in this contribution aspects of the CR primary proton spectrum are discussed, in the kinetic energy range 0.2 ÷ 200 GeV, in the neighbourhood of Earth, as have been measured by the Alpha Magnetic Spectrometer during space shuttle flight STS-91. The energy range between $10^5 \div 10^8$ TeV of the CR is closed by the expected energy of protons at ULHC, but the flux is irrelevant for degradation effects in detectors to be considered.

In this work we discuss how to prepare some possible detectors – only silicon option being considered, for the new era of HEP challenges, because bulk displacement damage in the detector, consequence of irradiation, produces effects at the device level that limit their long time utilisation, increasing the leakage current and the depletion voltage up to breakdown, and thus affects the lifetime of detector systems.

The main parameters of LHC and of future colliders, relevant for the present investigation, are presented in Table 1. For ULHC, estimates of some physical quantities are based on the linear extrapolation of known results, but surprises are not excluded at ultra high energies.

## 2. Interactions of Particles in the Detector and the of Kinetics of Defects

The nuclear interaction between the incident particle and the lattice nuclei produces bulk defects: the recoil nucleus(i) produced in the interaction are dis-placed from lattice positions into interstitials, their initial positions being vacant. Then, the primary knock-on nucleus, if its energy is large enough, could produce new displacements and the process continues as long as the ener-gy of the colliding nucleus is higher than the threshold for atomic displa-cements. The physical quantity characterising the process is the concentration of primary defects produced per unit of fluence of incident particles (CPD).

The values of CPD corresponding to the expected <$E_T$> energies for the spectra after primary interactions at future colliders are in the vicinity of the asymptotic high energy limit, where:

$$CPD^{asymp}(E) \cong \frac{N}{2E_d} \sum_i L(E_{Ri})\sigma_i < \frac{N}{2E_d} L^{\max}(E_R)\sigma_{tot}$$

and have approximately the values in the same domain. Here $E$ is the kinetic energy of the incident particle, $N$ is the atomic density of the target element, $E_d$ is the threshold energy for displacements (function on the orientation of the target lattice), $E_{Ri}$ - the recoil energy of the residual nucleus produced in interaction $i$, and $\sigma_{i(tot)}$ - the cross section of the interaction between the incident particle and the nucleus of the lattice for the process or mechanism $(i)$, responsible in defect production or the total cross-section respectively.

CPD is not proportional to the modifications of material parameters after irradiation, due to the subsequent interactions of vacancies and interstitials and with other defects and impurities (only phosphorus, carbon and oxygen are considered as pre-existent) in the lattice.

The formation and time evolution of complex defects, associations of primary defects or of primary defects and impurities are studied in accord with the model developed by the authors; see for example [2] and previously author's papers cited therein.

The primary point defects considered here are silicon self interstitials and vacancies: "classic" vacancies and fourfold coordinated defects - $Si_{FFCD}$; new defect predicted by Goedecker and co-workers [3]. The characteristics of the $Si_{FFCD}$ defect have been determined indirectly by the authors [4]. So, $Si \xrightarrow{G} (V + Si_{FFCD}) + I$, where the rate $G$ is a sum of contributions from thermal generation and irradiation. The integral irradiation rate is a convolution between CPD and the flux of particles. In the irradiation rate, the information about the identity and characteristics of primary particles that produce degra-dation is lost. The phenomena occurring in the material after primary defect production are considered in the frame of the theory of diffusion limited reactions. The complete list of reactions considered was done in Ref. [4]. The model predicts absolute values for microscopic quantities – concentrations of defects and the characteristics of the device, without free parameters. A good accord with available experimental data was obtained [4].

## 3. Predictions of the Behaviour of Detectors

The experimental results and model calculation for step by step irradiation at high rates of generation of defects, up to fluences expected at new colliders, indicate a very good agreement see [5], but the degradation is not clearly correlated with oxygen concentration in silicon or with the resistivity of the starting material. Consequently, for these experimental conditions, it is not possible to make a choice between FZ and DOFZ materials. In Figs. 1 and 2, the time dependence of the leakage current and of effective carrier concentration of for FZ and DOFZ materials, in conditions of continuous irradiation at rates between those corresponding to cosmic protons up to new colliders are presented. The rates of generation of defects are about 200 pairs/cm$^3$ for cosmic particles, about $7\times10^8$ pairs/cm$^3$ for LHC, and roughly 10 times higher for SLHC and VLHC.

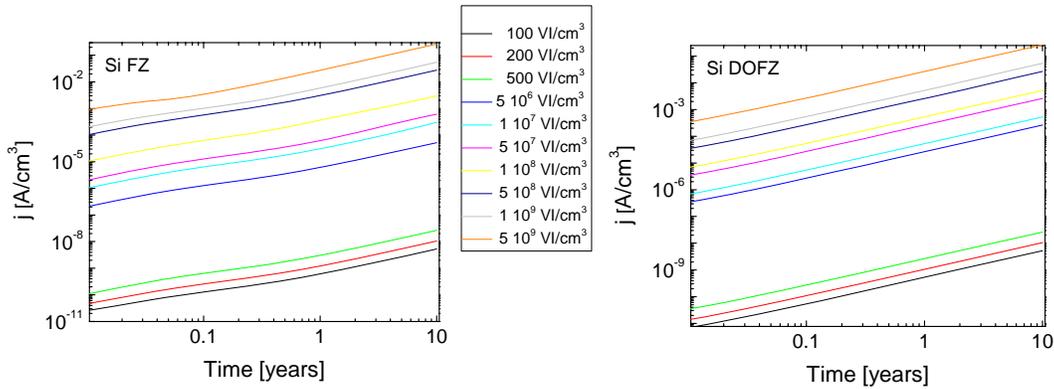

Figure 1.
Time dependence of the volume density of the leakage current for FZ and DOFZ silicon.

For leakage current, an approximately linear dependence of time is obtained in log-log scale, FZ materials presenting a lower slope in comparison with DOFZ ones. The produced degradation of silicon detectors scales with luminosity, and is roughly independent on particle spectra. If silicon detectors are exposed at primary protons fluxes from CR, the degradations are minor, and for these rates of generation of defects, DOFZ material is more adequate. For higher rates, both materials present inversion. In Fig. 2, the rates corresponding to the beginning of inversion are also presented.

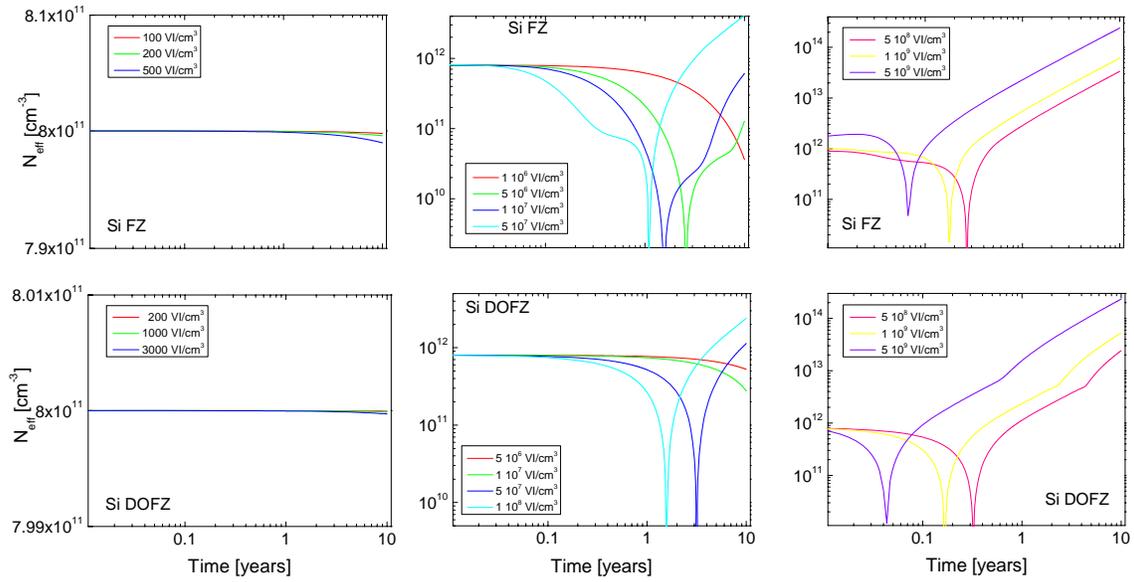

Figure 2.
Time dependence of $N_{eff}$ for detectors from FZ and DOFZ Si, for different rates of irradiation

## 4. Conclusions and Summary

The present model is able to reproduce main features of available experimental data of FZ and DOFZ Si and permits to understand the mechanisms of degradation. At high rates of generation of defects, the contribution of primary defects is important and must be considered. No major differences in degradation between FZ and DOFZ technologies are obtained and from this point of view a decision is not possible. Because the processes are temperature dependent, in order to diminish these macroscopic effects, the decrease of the temperature is strongly required.